\documentclass[11pt]{article} 

\usepackage[utf8]{inputenc} 

\usepackage{geometry} 
\usepackage{graphicx} 

\usepackage{amsmath,amsfonts,amsthm}

\usepackage{booktabs} 
\usepackage{array} 
\usepackage{paralist} 
\usepackage{verbatim} 
\usepackage{subfig} 

\usepackage{fancyhdr} 
\usepackage{sectsty}
\allsectionsfont{\sffamily\mdseries\upshape} 
\newtheorem{Def}{Definition}

\usepackage[nottoc,notlof,notlot]{tocbibind} 
\usepackage[titles,subfigure]{tocloft} 

\numberwithin{equation}{section}

\title{Vertex Operators, $\mathbb{C}^3$ Curve, and Topological Vertex}

\date{} 

\begin{document}
\maketitle
\centerline{Jian-feng Wu $^{1}$ and Jie Yang$^{2,3}$}\vspace{.1in}
\centerline{\emph{$^1$\hspace{-0.1cm}
Institute of Theoretical Physics, Department of applied mathematics}}
\centerline{\emph{\hspace{1cm}and physics,  Beijing University of Technology,
 Beijing, 100124, China}}
\centerline{ muchen.wu@gmail.com}
\vspace{.1in}
\centerline{$^2$\hspace{-.1cm}\emph{Beijing Center for Mathematics and Information Interdisciplinary Sciences}}
\vspace{.1in}
\centerline{$^3$\hspace{-.1cm} \emph{School of Mathematical Sciences, Capital Normal University, Beijing, 100048, China}}
\centerline{ yang9602@gmail.com}
\abstract
In this article, we prove the conjecture that Kodaira-Spencer theory for the topological vertex is a free fermion theory. By dividing the $\mathbb{C}^3$ curve into core and asymptotic regions and using Boson-Fermion correspondence, we construct a generic three-leg correlation function which reformulates the topological vertex in a vertex operator approach. We propose a conjecture of the correlation function identity which in a degenerate case becomes Zhou\rq{}s identity for a Hopf link.

\section{Introduction}
It has been proposed that the Chern-Simons theory of a gauge group $U(N)$ in the large $N$ limit is dual to A-model topological string theory \cite{gopakumar1999gauge}. \cite{ooguri2000knot} provided a brane configuration of the knot of Chern-Simons theory. \cite{witten1989quantum} discovered a quantum structure of Chern-Simons theory and from the well-known Wess-Zumino-Witten (WZW) model it even discovered the deeper relation between knot invariants in Chern-Simons theory with $SU(2)$ gauge group and characters of WZW model. Later the gauge group has been generalized to $U(N)$ \cite{morton2003homfly}. In the large $N$ limit some knot invariants such as the loop and the Hopf link are shown to be directly related to symmetric functions such as Schur and skew Schur functions. In \cite{Okounkov:2003sp} the authors discovered some interesting vertex structure for some geometrical and physical invariants such as the Donaldson-Thomas invariants of ${\mathbb C}^3$ or in physical language, BPS invariants of D0-D6 branes on ${\mathbb C}^3$. From statistical point of view it is the partition function of a crystal melting model. \cite{top_vertex} extended this structure to a general case where there are asymptotic boundaries of those crystals and related the partition function to a topological vertex. They managed to build up this connection because of Zhou's identity \cite{zhou2003conjecture}.

In \cite{ADKMV}, the authors achieved a B-model approach to the topological vertex based on the observation of the mirror curve of $\mathbb{C}^3$ and the related symmetries. However, an explicit correspondence between A-model and B-model is still an open question.

We try to find a more obvious relation between A- and B-model in this article. In our approach, the curve of $\mathbb{C}^3$ is essential. In A-model description, $\mathbb{C}^3$ could be seen as a cotangent bundle $T^*S^3$ with a single $S^3$ as the base. However, as the CS/WZW correspondence \cite{witten1989quantum} saying, topological invariants in A-model becomes correlation functions in B-model, where there is an modular $S$ transformation inserted between bra and ket states. This correspondence strongly implies the mirror curve of $\mathbb{C}^3$ can not be expressed simply in one coordinate chart. We need at least two coordinate charts , one being related to another by $S$ transformation. Thus a complete B-model mirror curve of $\mathbb{C}^3$ has an asymptotic region (near infinity where the bra state is inserted in) and a core region (near origin where the ket state is inserted in) with point 1 the fix point. This means the A-model theory is a union of CFTs in two regions with a defect inserted at point 1. Surprisingly, if we introduce an excitation at point 1, the Hamiltonian blows up the excitation and forms a distribution corresponding to the representation of the excitation. This is very much like the so-called projective representation of affine algebra as in \cite{okounkov2001infinite}. Also in \cite{Iqbal2009}, this structure had been introduced without proof.

The CFT considered at hand is a Kodaira-Spencer theory as explained in \cite{ADKMV}, which by definition is a bosonic theory, equipped with a broken $\mathcal{W}_{1+\infty}$ symmetry. A conjecture also was proposed in \cite{ADKMV} that the corresponding fermionic theory is a free fermion theory. We prove in this article that for the topological vertex case, where the unbroken $W$ symmetry is $W^3_0$, this conjecture is true. For other cases, for example, $W^4$, if one would like to keep the integrable structure, the corresponding fermionic theory is still a free one. It is compatible with Dijkgraaf\rq{}s work \cite{dijkgraaf1997chiral} two decades ago.

Free fermion has been used in many research areas of physics. In \cite{douglas1995conformal} two-dimensional Yang-Mills theory of $U(N)$ gauge group the Vandermonde of group measure implies there is a fermionic structure. In two dimensions, due to Boson-Fermion correspondence, vertex operator is a very useful tool. In B-model \cite{ADKMV} provided a beautiful explanation of a B-brane insertion as a fermion field and symmetries of the Riemann surface as the sources of transition function of sections of fiber bundles of fermionic fields. Because the duality between A-model and B-model, we would expect a similar structure in the A-model side. In this paper we aim to discover this structure and approach this topic via a vertex operator formalism.

The structure of this paper is following. In sec. \ref{sec:notation} we clarify some notation to be used in this paper and make some preparation. In sec. \ref{sec:vertex} we provide a generating function for the vertex operator. In sec. \ref{sec:curve} we obtain the fermionic expression for $W^n_0$ and prove the free fermion conjecture for $W^3_0$. We also examine the curve of $\mathbb{C}^3$ in different regions and related symplectic transformations. In sec. \ref{sec:ToV} we solve Hamiltonian equations for two different coordinate charts and obtain wave functions. We construct the correlation function with three generic representations inserted at three points: 0,1,$\infty$ in a single patch. The cyclic symmetry of the vertex becomes a conjecture of an identity for the correlation function. In the limitation situation, this correlation function identity becomes Zhou\rq{}s identity of Hopf link. In sec. \ref{sec:conclusion} we point out the future working direction.

\section{Notations and Preliminaries}\label{sec:notation}

A partition $\lambda$ is any sequence $\lambda = (\lambda_1\,, \lambda_2\,, \lambda_3\,, \cdots)$ of non-negative integers in weakly decreasing order: \[\lambda_1\geq\lambda_2\geq\lambda_3\geq \cdots .\] The diagram of a partition $\lambda$ may be formally defined as the set of points $(i, j)\in {\mathbb Z}^2$ such that $1\leq j\leq \lambda_i$. More often it is convenient to replace the points by squares. It is also called a Young diagram.  The conjugate of a partition $\lambda$ is denoted by $\lambda^t$ whose diagram is obtained by reflection in the main diagonal of $\lambda$.

A Schur function $s_{\lambda}(z_i)$ is a symmetric function of $z_i\,,i=1,2,\cdots,\infty$'s and labeled by partition $\lambda$. Especially, when \[z_i = q^{\rho_i} =q^{-i+\frac{1}{2}}\] there is a very useful product formula for $s_{\lambda}$
\begin{equation}
\label{eq:hookschur}
s_{\lambda}(q^{-\rho}) = \frac{q^{||\lambda^t||/2}}{\prod_{(i,j)\in\lambda} 1-q^{h(i,j)}}\,,\end{equation} 
where $||\lambda^t|| = \sum_i{(\lambda^t_i)^2}$ and  $h(i,j)$ is the hook length of square $(i,j)$. Sometimes it is more useful to represent the hook length as $h(i,j) = a(i,j)+l(i,j)+1$ where $a(i,j)$ and $l(i,j)$ are arm-length and leg-length respectively and 
\begin{eqnarray*}
a(i,j) = \lambda_i - j \,, \quad l(i,j) = \lambda^t_j -i .
\end{eqnarray*}
\subsection{Zhou\rq{}s Hopf link identity}
 The Hopf link is defined by
\begin{equation}
W_{\lambda\mu}=W_{\lambda}\,s_{\mu}(q^{\lambda+\rho}),
\end{equation}
where
\begin{equation*}
W_{\lambda}=(-1)^{\lambda}q^{\kappa_\lambda/2}\,s_{\lambda}(q^{-\rho})=s_\lambda(q^\rho),
\end{equation*} 
$\rho=-\frac{1}{2},-\frac{3}{2},-\frac{5}{2},\cdots$, and $q=e^{g_s}$. According to the duality between Chern-Simons and A-model topological string theory,
$g_s$ is the string coupling constant. $\kappa_{\mu}$ is the "energy" of the representation $\mu$ and \[\kappa_\mu = \sum_i{\mu_i^2}-\sum_j{\mu^t_j}^2=\sum_{i\leq \ell(\mu)}\left((\mu+\rho)_i^2-\rho_i^2\right) \]
where $\ell(\mu)$ is the length of the partition $\mu$, that is $\mu_1^t$. Zhou's Hopf link identity can be written as
\begin{eqnarray}
q^{\kappa_{\mu^t}/2}s_{\lambda}(q^{-\rho})s_{\mu^t}(q^{-\lambda^t-\rho})=\sum_{\eta}s_{\lambda/\eta}(q^{-\rho})s_{\mu/\eta}(q^{-\rho})\,,
\end{eqnarray}
where $s_{\lambda/\eta}(q^{-\rho})$ is the skew Schur function, see \cite{Macdonald1999} for a detailed definition.
For a special case $\lambda=\phi$ it gives rise to an interesting identity \footnote{$\phi$ denotes the empty Young diagram}:
\begin{equation}\label{singleschur}
s_{\mu}(q^{-\rho}) = q^{\kappa_{\mu^t}/2}s_{\mu^t}(q^{-\rho})\,. 
\end{equation}
With the help of the useful identity \footnote{\cite{Okounkov:2003sp} and \cite{zhou2003curve} obtained a different formula \[s_{\mu/\eta}(q^{-\nu-\rho}) = (-)^{|\mu|+|\eta|}s_{\mu^t/\eta^t}(q^{\nu+\rho})\,,\] which is not correct. In \cite{zhou2003curve} there was only a minor typo in the last line in the derivation of eq. (28). In \cite{Okounkov:2003sp} there was no derivation. We provide an independent derivation in App. \ref{app:id}. We thank Professor Guo-ce Xin for pointing out the problem for us.}\cite{Okounkov:2003sp, zhou2003curve}:
\[s_{\mu/\eta}(q^{-\nu-\rho}) = (-)^{|\mu|+|\eta|}s_{\mu^t/\eta^t}(q^{\nu^t+\rho})\]
we obtain
\begin{eqnarray}
W_{\lambda\mu} = (-)^{|\lambda|+|\mu|}q^{\kappa_\lambda/2+\kappa_\mu/2}\sum_{\eta}
s_{\lambda/\eta}(q^{-\rho})s_{\mu/\eta}(q^{-\rho}).
\end{eqnarray}
\cite{zhou2003conjecture} provided a mathematical proof of this Hopf link identity.  Since this identity has a lot of applications in both mathematics and physics, see, e.g. \cite{top_vertex, zhou2013explicit}, we expect to uncover its origin from a physical point of view. However, a concrete physical proof is still an open problem for us.

\section{Vertex Operators and Generating Functions}\label{sec:vertex}
Let us first introduce some basic ingredients of conformal field theory (CFT) of holomorphic boson and also chiral fermion fields.
\subsection{Bosonization and Fermionization}
For a fermion-antifermion system defined on a complex plane, we have the following chiral fermion fields (Neuve-Schwarz fermions):
\begin{eqnarray}
\psi(z) = \sum_{r\in\mathbb{Z}+\frac{1}{2}}\psi_r z^{-r-1/2}\,,  \\\nonumber
\psi^*(z) = \sum_{s\in\mathbb{Z}+\frac{1}{2}}\psi^*_s z^{-s-1/2}.
\end{eqnarray}
They have the following operator product expansions (OPEs):
\begin{eqnarray}\label{ChiralOPE}
\psi(z)\psi^*(z') &=&\frac{1}{z-z'}:\psi(z)\psi^*(z'):+\cdots\, \\\nonumber
\psi^*(z)\psi(z') &=& \frac{1}{z-z'}:\psi^*(z)\psi(z'): +\cdots\,\,,
\end{eqnarray}
and also the anti-commutation relations:
\begin{equation}
\{\psi_n , \psi^*_m\} = \delta_{n+m,0}\,, \quad \text{others} = 0\,,\quad \quad
(\psi_n)^*=\psi^*_{-n}\,.
\end{equation}

In another side, the holomorphic bosonic field  $\varphi(z)$ is given as following 
\begin{eqnarray}\label{phi}
\varphi(z) &=& q_0+ p_0\ln z+ \sum_{n\neq 0}\frac{a_n}{-n}z^{-n}, \\\nonumber
 [a_n, a_m] &=& n\delta_{n+m,0}\,\,,\, [p_0,q_0]=1\\
\varphi(z) \varphi(w) &=& \ln(z-w):\varphi(z) \varphi(w): \,,
\end{eqnarray} 
If $\bar{\varphi}$ denotes the corresponding anti-holomorphic bosonic field then a free bosonic field $\varphi(z,\bar{z})$ can be constructed as:
\[\varphi(z,\bar{z}) = \varphi(z)-\bar{\varphi}(\bar{z})\,.\]

In this article, we only make the holomorphic part of the bosonic field dynamic and leave the anti-holomorphic part non-dynamic. While this boson is called a chiral boson and the corresponding vertex operators are called chiral vertex operators. A chiral vertex operator may be written as 
\[V_\alpha(z) = e^{\alpha\varphi(z)},\] its conjugation is
\[(V_{\alpha}(z))^* = e^{-\alpha\varphi(z^*)}= V_{-\alpha}(z^*).\]
However, we just consider the case that $\alpha=1$ and denote
\begin{eqnarray}
V(z) = e^{\varphi(z)}\,\,,\,\,
V^*(z) = e^{-\varphi(z)}.
\end{eqnarray}
From modes expansion and Heisenberg algebra (\ref{phi}), we can calculate the OPEs of $V(z)$ and $V^{*}(z')$:
\begin{eqnarray}
\label{Vself}
:V(z)::V(z'):  &=&   :VV(z\rq{}):(z-z')+reg.\\\nonumber
:V^{*}(z)::V^{*}(z'):&=&:V^{*}V^{*}(z\rq{}):(z-z')+reg.\\\label{eq:comm_fin}
:V^{*}(z)::V(z'):&=&:V^{*}V(z'):\frac{1}{(z-z')}-\partial{\varphi(z')}+reg.\\\nonumber
:V(z)::V^{*}(z'):&=&:V V^{*}(z'):\frac{1}{(z-z')}+\partial{\varphi(z')}+reg.,
\end{eqnarray}
here $reg.$ means regular terms.
The singular parts of these OPEs are the same as those of chiral fermions in eq.(\ref{ChiralOPE}). However chiral fermions have no self-contractions they are not completely the same as $V$ and $V^*$. Nevertheless according to Pauli's exclusive principle, namely the fermionic statistics, the correlation function of fermions is related to Slater determinant
\begin{eqnarray}
\langle vac|\prod_{i=1}^N \psi(z_i)\prod_{j=1}^N\psi^*(w_j) |vac\rangle& = &\langle 0|\prod_{i=1}^N V(z_i) \prod_{j=1}^NV^*(w_j)|0\rangle \nonumber \\
={\rm Det}(\frac{1}{z_i - w_j})&=&\frac{\prod_{i<j}(z_i - z_j)(w_i-w_j)}{\prod_{i,j =1}^N(z_i-w_j)} \,.
\end{eqnarray}
It reminds us the miraculous boson/fermion correspondence.
Therefore during the calculation of the correlation function, we can replace all fermionic fields by bosonic vertex operators such that
\begin{equation}
\psi(z)\sim V(z) = e^{\varphi(z)}\,,\quad \psi^*(z)\sim V^*(z) = e^{-\varphi(z)}\,.
\end{equation}
 However, we need to bear it in mind that chiral fermions are not exactly these vertex operators because of the different self OPEs. Secondly there could be different numbers of $V$ and $V^*$ in the correlation function by carefully choosing the charges of bra and ket vacua. But if there are different number of $\psi$ and $\psi^*$, the correlation function is automatically vanishing.
 
In another way, since both fermionic and bosonic theories have the same $U(1)$ symmetry, the charge is measured by the number of $\partial \varphi(z)$ in the bosonic theory and $\psi\psi^*(z)$ in the fermionic theory. Hence we have the fermionization as follows:
\begin{equation}
\partial\varphi(z)=(\psi\psi^*)(z)\,,
\end{equation}
in terms of the modes expansion that is
\begin{equation*}
  a_n =\sum_{r\in{\mathbb Z}+1/2} :\psi_{n-r}\psi^*_r:\,.
\end{equation*}

 \subsection{Generating Functions} 
In the following we denote $V_+$ and $V_-$ as the positive and negative modes part of $e^\varphi$ and $V_+^*$ and $V_-^*$ as the corresponding part of $e^{-\varphi}$, that is,
\begin{eqnarray}
  \label{eq:Vertex}
  V_+(z)=\exp\left\{\sum_{n>0}\frac{a_n}{-n}z^{-n}\right\},\quad V_-(z)=\exp\left\{\sum_{n>0}\frac{a_{-n}}{n}z^{n}\right\},\\\nonumber
   V_+^{*}(z)=\exp\left\{\sum_{n>0}\frac{a_n}{n}z^{-n}\right\},\quad V_-^*(z)=\exp\left\{\sum_{n>0}\frac{a_{-n}}{-n}z^{n}\right\}\,.
\end{eqnarray}
They form four types of generating functions of Schur functions, namely
\begin{eqnarray}
  \prod_i V_-(z_i)&=&\sum_\lambda s_\lambda(z_i)s_\lambda(a_-),\\
  \prod_i V_+^*(z_i)&=&\sum_\lambda s_\lambda(z_i^{-1})s_\lambda(a_+),\\
  \prod_i V^*_-(z_i)&=&\sum_\lambda(-1)^{|\lambda|}s_\lambda(z_i)s_{\lambda^t}(a_-),\\
  \prod_i V_+(z_i)&=&\sum_\lambda(-1)^{|\lambda|}s_\lambda(z_i^{-1})s_{\lambda^t}(a_+).
\end{eqnarray}
We will use these generating functions frequently in our calculation. 
They also can be deduced from four basic generating functions such that:
\begin{eqnarray}\label{EHgenerating}
V_-(z) &=& \sum_{r\in\mathbb{Z}^+} s_{(r)}(z)s_{(r)}(a_-)\\\nonumber
V_-^*(z)&=&\sum_{r\in\mathbb{Z}^+}(-)^r s_{(r)}(z)s_{(1^r)}(a_-)\\\nonumber
V_+(z) &=& \sum_{r\in\mathbb{Z}^+}(-)^r s_{(r)}(1/z)s_{(1^r)}(a_+)\\\nonumber
V_-^*(z)&=&\sum_{r\in\mathbb{Z}^+} s_{(r)}(1/z)s_{(r)}(a_+)\,,
\end{eqnarray}
where $(r)$ ($(1^r)$) denotes a length-$r$ horizontal (vertical) Young diagram. Actually, the Schur polynomials of the horizontal and vertical Young diagrams are the same as the complete (homogeneous) and elementary symmetric polynomials respectively, that is,
\[s_{(r)}(z) = h_r(z),\quad s_{(1^r)} =e_r(z)\,.\]

 \subsection{Fermionic Vacua and Maya/Young Correspondence}
The vacuum of free fermion theory corresponds to a filled Dirac sea. Firstly, for the ket vacuum, we denote $|\Omega\rangle$ as the 'fake' vacuum of the theory, which is annihilated by all modes of $\psi$ but not $\psi^*$. Since $\psi^*$ is the anti-particle field of $\psi$, the positive modes of $\psi^*$ should be understood as creation operators of anti-particles. The 'real' (physical) ket vacuum should have a natural Dirac sea structure and is denoted by $|vac\rangle$ and defined as
 \begin{equation}|vac\rangle = \psi_{1/2}^*\psi_{3/2}^*\psi_{5/2}^*\cdots|\Omega\rangle\,.\end{equation}
The bra vacuum could be defined by conjugation of ket vacuum
\[\langle vac| = \langle\Omega |\cdots\psi_{-5/2}\psi_{-3/2}\psi_{-1/2}\,.\]
A unitary excitation on the ket vacuum $|vac\rangle$ always contains pairs of particle-anti-particle. To track the sign of the excited state, we would better define an excited state as follows
\begin{equation}\label{excitedstate}
(-)^{\sum_i^n s_i-1/2}\prod_{i=1}^n \psi_{-r_i}\psi^*_{-s_i}|vac\rangle\,,
\end{equation}
where the subscripts $r_i, s_i$ are positive half integers ($\mathbb{Z}_{>0} -1/2$).
 
Since the choice of particle or anti-particle is arbitrary we could choose the 'fake' vacuum $|\Omega'\rangle$ as the one annihilated by all modes of $\psi^*$. Therefore the definition of bra and ket vacua should be changed as
\begin{eqnarray*}&&|vac'\rangle = \psi_{1/2}\psi_{3/2}\psi_{5/2}\cdots|\Omega'\rangle\\
&&\langle vac'| = \langle \Omega' |\cdots\psi^*_{-5/2}\psi^*_{-3/2}\psi^*_{-1/2}\,.
\end{eqnarray*} 
Similarly we can define corresponding excited states. 

There is a transformation switching these two choices of vacua which is called  an involution and denoted by $\omega$. It acts on the states and operators as follows
\begin{eqnarray}\label{Involution}
\omega: |\Omega\rangle \rightarrow |\Omega'\rangle \,, \quad 
\psi^*_n\rightarrow (-)^n\psi_n \,, \quad \psi_n\rightarrow (-)^n\psi^*_n\,.
\end{eqnarray}

\begin{figure}[h]
  \centering
  \includegraphics[width=0.8\textwidth]{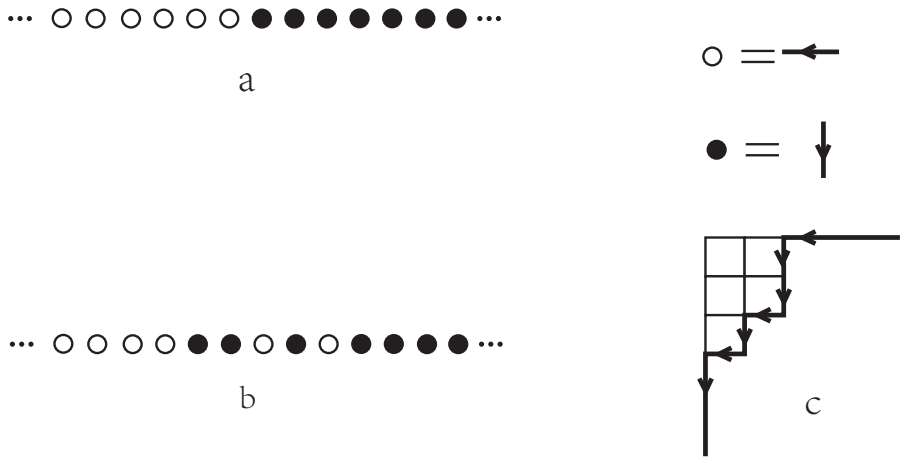}
  \caption{a. The Maya diagram for the vacuum, b. the Maya diagram for the excited state $(r_1=3/2, r_2=1/2, s_1=5/2, s_2=1/2)$, c. the corresponding Young diagram of b.}
  \label{fig:Maya}
\end{figure}

For states defined by eq. (\ref{excitedstate}), we can use Maya diagram to demonstrate the excitations. For example, the Maya diagram corresponding to the vacuum is shown in fig. \ref{fig:Maya}a. An excited state can be obtained by exchanging certain white and black dots of the vacuum Maya diagram. For example Fig. \ref{fig:Maya}b. shows an excited state denoted by $(r_1=3/2, r_2=1/2, s_1=5/2, s_2=1/2)$. 

There is an amazing correspondence between Maya diagrams and Young diagrams. For each white/black dot, we assign a unit leftward/downward line segment connected end to end. Hence a Maya diagram corresponds to a unique Young diagram in this setup. Thus the subscripts $r_i, s_i$ are Frobenius coordinates defining the Young diagram. Fig. \ref{fig:Maya}c. shows the corresponding Young diagram of Maya diagram in Fig. \ref{fig:Maya}b. which is $\{2,2,1\}$. As shown in \cite{Jimbo1983}, the excitation states defined in this way are Schur states in fermionic operator representation.

\section{Curve driving Patch-Shifting}\label{sec:curve}

Previously, we have considered the definition of bosonic and fermionic theory on the complex plane. However, we are more interested in theory on special Riemann surface with punctures and also defects \cite{drukker2011virtue}. By special we mean the Riemann surface is obtained by gluing various regions, with defects inserted at fix points.   

To obtain Riemann Surface with more complicated topology, we need to define theories on tubes and pants. A conformal field theory defined on a tube is a boundary CFT (BCFT) \cite{cardy2004boundary} with two boundary conditions. While on a pants, the corresponding CFT is a BCFT with three boundary conditions. 

However, the problem at hand differs from  a BCFT problem because the theory is not defined on a simple Riemann surface but with defect inserted. Precisely, we have a core region in the toric structure of $\mathbb{C}^3$ and three asymptotic regions which are local patches and  can be transited among themselves. A defect is inserted at point $z=1$\footnote{The defect could be anywhere on the complex plane, however, by global conformal transformation, it can be fixed at point 1 without loss of generality.}. Therefore a BCFT analysis may fail in this case.

As shown in \cite{ADKMV}, the Riemann surface corresponding to toric Calabi-Yau could be defined patch by patch and there are some symplectic transformations of coordinates of local patches. A simpler case for two-punctured theory can be obtained by joining two patches together.  That is to say, if we define a theory on a local patch associated with one of these two punctures, then another theory on another local patch, these two theories can be related to each other by patch-shifting transformation, which is a symplectic transformation.
The cut and join operation should preserve the symplectic transformation from one patch to another.

Symplectic transformation is an area-preserving operation which is compatible with the measure on Riemann surface.

\subsection{The $\mathcal{W}_{1+\infty}$ Algebra}

Now for an infinite cylinder, the symplectic form is \[\Omega = d x\wedge d p\,, \] hence the symplectic transformation of $x$ is
\begin{equation}\label{symp}x\rightarrow x+\epsilon(x) = x+f(p) = x +\sum_{n=0}^{\infty} f_n p^n\,.\end{equation}

The transformation of a quantum chiral scalar field associated with the change of the local coordinate $\delta x = \epsilon(x)$ is implemented
by the operator \begin{equation}
\oint T(x)\epsilon(x)\frac{d x}{2\pi i},
\end{equation}
where the stress tensor for chiral boson theory is \[T(x) = \frac{1}{2} [\partial\varphi\partial\varphi](x)\,.\]
The observables of the chiral bosonic theory correspond to variations of the complex structure
at infinity. On each patch this is described by the modes of a chiral boson $\varphi(x)$, defined by \[\partial \varphi(x) =p(x)\,.\] 

Now the symplectic transformation has an operator expression, namely
\begin{eqnarray}
\oint T(x)\epsilon(x)\frac{dx}{2\pi i}  &=& \lim_{x\rq{} \rightarrow x}\oint \frac{1}{4\pi i} [\partial\varphi\partial\varphi](x)\sum_{n\geq 0}f_n[(\partial\varphi)^n](x\rq{}) d x \\\nonumber&=&\frac{1}{2}\sum_{n\geq 0}(n+2)!f_n W_0^{n+2}(x)+ terms \,\,involve\,\, (\partial^3\varphi)\,.\label{eq:symp}
\end{eqnarray}
 $W_0^{n}$ is the zero mode of free (non-interacting) $\mathcal{W}^n$ transformation which is defined by
\begin{eqnarray}
W^n(z) &=& \frac{1}{n!}(z\partial_z \varphi(z))^n\\
W^n_m &=& \oint \frac{1}{ 2\pi i z}\frac{1}{n!} z^{-m+n}(\partial_{z}\varphi(z))^n\\\nonumber&=& \frac{1}{n!}\sum_{k_i=-\infty }^{\infty}\delta\left((\sum_{i=1}^n k_i)-m\right)\left(:a_{-k_1}a_{-k_2}\cdots a_{-k_n}:\right) \,,
\end{eqnarray} 
up to some constant ground energy due to normal ordering. In the derivation, it is useful to apply the coordinate transformation from the cylindrical coordinates to the complex plane ones $z = e^{x}$. Thus \[z ^{-1}d z =  d x, \partial_x = z\partial_z\,, \partial \varphi(x) = z\partial_z\varphi(z) =\sum_n a_n z^n, \,.\]

The appearance of a term containing $\partial^3\varphi$ in eq. (\ref{eq:symp}) reflects the non-associativity (and also non-commutation) of operator expansion product meaning
\[[A[BC]](z)\neq [[AB]C](z)\,.\]
This is crucial in the derivation of operator formalism of a given integral formula. Next we will use  $W^3_0$ and $W^4_0$ as two examples to explain it.

Firstly, we have 
\begin{eqnarray}\label{W30}
&&[\partial_x\varphi\partial_x\varphi\partial_x\varphi](x) = z^3[(\partial_z\varphi)^3](z) \\\nonumber &=&\frac{z^3}{2\pi i} \oint_z \frac{1}{z\rq{}-z}\partial_{z\rq{}}\varphi(z\rq{})[\partial_{z}\varphi\partial_z\varphi](z)d z\rq{}\end{eqnarray} while
\begin{eqnarray}\label{W301}
&&[(\partial_x\varphi)^3](x)+[\partial_x^3\varphi](x)\\\nonumber
&=& \frac{z^3}{2\pi i}\oint_z\frac{1}{z\rq{}-z}[\partial_{z\rq{}}\varphi\partial_{z\rq{}}\varphi](z\rq{})\partial_z \varphi(z)dz\rq{}\,.
\end{eqnarray} Secondly, by using the modes expansion as we defined before, we have the bosonic operator formalism of (\ref{W30})  such that
\begin{eqnarray}\label{W3boson}W^3_0 &=&\frac{1}{6} \oint_x dx \frac{1}{2\pi i}[\partial_x\varphi\partial_x\varphi\partial_x\varphi](x) \\\nonumber&=& \frac{1}{2}\sum_{n,m>0}(a_{-n-m} a_m a_n + a_{-n}a_{-m}a_{n+m}) +\frac{1}{2}\sum_{n>0}a_0a_{-n}a_{n}+\frac{1}{6}a_0^3\,.\end{eqnarray}

In another way, the bosonic operator formalism of (\ref{W301}) gives rise to
\begin{eqnarray}
\widetilde{W}^3_0 &=&\frac{1}{6}\oint_x dx \frac{1}{2\pi i}  \left([(\partial_x\varphi)^3](x)+[\partial_x^3\varphi](x)\right) = W^3_0 + \frac{1}{3}a_0\,.
\end{eqnarray}
Here $a_0=p_0$ is the momentum of the center of mass. Hence the last term  is not an important correction for the spectrum of $W^3_0$. But it is fascinating for us that $\frac{1}{3}a_0$ also appears in the fermionic side which seems to provide further evidence for Boson-Fermion correspondence and we will proceed to that point later.

The non-associative property leads to a severe problem, that is
\[[W^n,W^m]\neq 0, \,\,for\,\,n\neq m\,.\] This is a displeasing result since we  would expect a $\mathcal{W}_{1+\infty}$ symmetry to generate the integrability, which means there are infinite many conserved currents commuting with each other. Another problem is that it is difficult to construct the exact form of $W^n$ in terms of bosonic fields. However, the difference of $W^3_0$ and $\widetilde{W}^3_0$ reveals a simple fact: the difference is just a total derivative! Actually, $\widetilde{W}_0^3$ rather than $W^3_0$ is the one included in the $\mathcal{W}_{1+\infty}$ algebra.

If we proceed to the fourth $W$ generator, there are three different choices. An arduous way to find the correct one is to apply the commutation on those choices with the lower order $W$ generators to check which one gives rise to zero simultaneously. However, it turns out to be simpler to approach this problem from the fermionic picture and we will elaborate it next.

Up to a total derivative, we see that $W^3$ is the same as $\widetilde{W}^3$. Moreover we can use a $\partial$-cohomology definition of $\mathcal{W}_{1+\infty}$ algebra which means the algebra is closed upon modulo all total derivatives. Then we get a good definition of $W^n$ algebra in terms of the bosonic formalism
\[W^n(z) = \frac{1}{n!}(\partial\varphi)^n \,(mod\,\,\partial)\,.\] This observation was known long ago since Dijkgraaf\rq{}s paper \cite{dijkgraaf1997chiral}. However, the derivation is not exactly the same. Actually it is quite astonishing for us. Since we are only considering how free chiral boson CFT goes from one patch to another patch keeping some symplectic symmetries, then we obtain the chiral boson theory which turns out to be a Kodaira-Spencer-like one as the previous work \cite{dijkgraaf1997chiral} by Dijkgraaf.

\subsection{Fermionic Representation }
We want to check the non-associative property in a fermionic picture. Firstly we consider the $W^3_0$ case. Since there is no significant difference between $W^3_0$ and $\widetilde{W}^3_0$ it is sufficient to examine $W^3_0$.

From fermionization \[\partial_z\varphi(z) = \psi\psi^*(z)\,, \] we can write down the fermionic formalism of $W^3_0$ as
\begin{eqnarray}\label{Ferm30}
6 W^3(w) &=& \oint_z \frac{dz}{2\pi i (z-w)}[\psi\psi^*](z)[-\psi\partial_w\psi^*-\psi^*\partial_w\psi](w)\\\nonumber
&=& -\oint_z \left(\frac{\psi(z)\partial_w\psi^*(w)}{(z-w)^2}+\frac{\psi^*(z)\psi(w)}{(z-w)^3}\right)\frac{dz}{2\pi i}\\\nonumber&+&\oint_z \left(\frac{\psi^*(z)\partial_w\psi(w)}{(z-w)^2}+\frac{\psi(z)\psi^*(w)}{(z-w)^3}\right)\frac{dz}{2\pi i}\\\nonumber
&=& 2[\partial\psi^*\partial\psi](w)+\frac{1}{2}[\partial^2\psi \psi^*](w)-\frac{1}{2}[\partial^2\psi^* \psi](w)\,.
\end{eqnarray} To get the first equality, we used the well-known result that
\[[\partial\varphi\partial\varphi](w) = -[\psi\partial\psi^*+\psi^*\partial\psi](w)\,.\] Actually, it can be treated as the first generalization of Boson-Fermion correspondence.
  The fermionic modes expansion leads to an operator formalism of $W_0^3$, namely
\begin{equation}
W_0^3  = \frac{1}{2}\sum_{r\in\mathbb{Z}_{>0}-\frac12}\left(r^2+\frac{1}{12}\right)(\psi_{-r}\psi^*_r-\psi^*_{-r}\psi_r)\,.
\end{equation}

Similarly, $\widetilde{W}_0^3$ has a fermionic expression
\begin{equation}\label{tFerm03}
6\widetilde{W}^3(w) = \frac{3}{2}[\partial^2\psi \psi^*-\partial^2\psi^* \psi](w)\,,
\end{equation} and the operator formalism
\begin{equation}
\widetilde{W}_0^3  = \frac{1}{2}\sum_{r\in\mathbb{Z}^+-1/2}\left(r^2+\frac{3}{4}\right)(\psi_{-r}\psi^*_r-\psi^*_{-r}\psi_r) = W^3_0 +\frac{1}{3}a_0\,,
\end{equation} where we substituted \(a_0 =\sum_r :\psi_{-r}\psi^*_r:\). 

In the last equality of (\ref{Ferm30}), the first term could be rewritten as
\[2\partial\psi^*\partial\psi =\partial(\psi^*\partial\psi-\psi \partial\psi^*)-\psi^*\partial^2\psi+\psi\partial^2 \psi^*\,.\] The operator formalism perfectly matches with previous bosonic result. A byproduct of this result is the generalization of Boson-Fermion correspondence to higher derivatives. For example we have the second generalization formula:
\begin{eqnarray}
[\partial^2\psi \psi^* - \partial^2\psi^*\psi](z)&=&\frac{2}{3}[(\partial\varphi)^3+\partial^3\varphi](z)\,.
\end{eqnarray}

Secondly the OPE method could be easily generalized to $W^4_0$ case while there are three ways of multiplication
\[[\partial\varphi][(\partial\varphi)^3],\,\,[(\partial\varphi)^2][\partial\varphi)^2]\,,\,\,[(\partial\varphi)^3][\partial\varphi]\,.\]
Hence there are three kinds of fourth-level Boson-Fermion correspondence as follows:
\begin{eqnarray*}
[(\partial\varphi)^4]&=&\left([\frac{3}{2}\partial\psi\partial^2\psi^*+\frac{1}{6}\partial^3\psi \psi^*]+\{\psi\leftrightarrow\psi^*\}\right)\\ & & \quad + [2\partial\psi^*\partial\psi\psi\psi^*]\\
\,[(\partial\varphi)^2][\partial\varphi)^2]&=&[2\partial^3
\varphi\partial\varphi+(\partial\varphi)^4]\\
&=&\left([\frac{4}{3}\partial^3\psi \psi^*-\partial^2\psi\partial\psi^*]+
\{\psi\leftrightarrow\psi^*\}\right)\\ & & \quad -[2\partial\psi^*\partial\psi\psi\psi^*]\\
\,[(\partial\varphi)^3][\partial\varphi]&=& [3\partial^3\varphi\partial\varphi+ 3(\partial^2\varphi)^2+ (\partial\varphi)^4]\\
&=&\frac{5}{3}\left[\partial^3\psi\psi^*+ \partial^3\psi^*\psi\right]\\ & &\quad + [2\partial\psi^*\partial\psi\psi\psi^*]\,.
\end{eqnarray*} 
These results have not modulo total derivatives yet. It is quite difficult to write down the operator formalism. All of them contain four-fermion terms which in general will spoil the integrable structure. If we use the $\partial$-cohomology definition, these three are the same up to total derivatives and constant factors. In fermionic picture, the $W^4$\footnote{We do not distinguish $W$ and $\tilde{W}$ explicitly from now on. }  could be rewritten as
\begin{eqnarray}
W^4(w) &=& \frac{1}{4}\oint\frac{dz}{2\pi i (z-w)}W^3(z)\psi\psi^*(w)
\\\nonumber
&=&\frac{1}{24}\oint\frac{dz}{2\pi i (z-w)}\frac{3}{2}[\partial^2\psi\psi^*-\partial^2\psi^*\psi](z)[\psi\psi^*](w)
\\\nonumber
&=& \frac{1}{12}(\partial^3\psi\psi^*+\partial^3\psi^*\psi)(w)\,.
\end{eqnarray}
A recursive derivation shows that the generic $W^n$ can be expressed as
\begin{eqnarray}
W^n(z)= \frac{1/2}{(n-1)!}\left(\partial^{n-1}\psi\psi^*+(-)^n
\partial^{n-1}\psi^*\psi\right)\,.
\end{eqnarray}
Therefore $W^n_0$ have distinct expressions for odd and even $n$\rq{}s. For odd $n$, 
\begin{equation}
W^n_0 \propto \sum_{r\in\mathbb{Z}^+-1/2}(\text{even power polynomial of} \,\,r)(\psi_{-r}\psi^*_r-\psi^*_{-r}\psi_r)\,,
\end{equation}
and for even $n$
\begin{equation}
W^n_0 \propto \sum_{r\in\mathbb{Z}^+-1/2}(\text{odd power polynomial of} \,\,r)(\psi_{-r}\psi^*_r+\psi^*_{-r}\psi_r)\,\,.
\end{equation}
Especially, the exact form of $W^4_0$ gives rise to
\begin{equation}\label{W40F}
W^4_0 = \left(\frac{r^3}{6}+\frac{23 r}{24}\right)(\psi_{-r}\psi^*_r+\psi^*_{-r}\psi_r)\,.
\end{equation}

A bonus of this fermionic operator formalism is that it reveals the integrable structure explicitly inherited from free fermions. The reason is that an eigenvector of these $W$ operators is formed by those pair excitations of fermions (\ref{excitedstate}) above Dirac sea as discussed before.

 This argument actually provides a proof of a conjecture proposed in \cite{ADKMV} where they pointed out that the Kodaira-Spencer chiral bosonic CFT exactly act like free chiral fermions. We have a stronger result that for a chiral boson action involving  $(\partial\varphi)^n$ ($n$ arbitrary) interactions, the fundamental theory is a free fermion theory.
 
 Although the bosonic theory is in general quite difficult to deal with, the corresponding fermionic one is rather simple. Moreover the integrable structure is explicit. 

\subsection{Quantum Curve and Patch-Shifting}

Now we consider the quantum curve for $\mathbb{C}^3$, which dominants the behavior of patch-shifting. Actually, the curve under consideration has distinct representation as a core and an asymptotic part, which are related by an $S$ transformation.

In \cite{ADKMV, top_vertex, bouchard2011topological, hori2000mirror} and \cite{2012arXiv1207.0598Z}, $\mathbb{C}^3$ has a mirror manifold defined by the algebraic equation
\begin{equation}zw-e^p-e^x-1=0\,,\end{equation}
the core curve of $\mathbb{C}^3$ is understood as
\begin{equation}\label{C3qcurve}
e^{p} + e^{x} + 1 =0\,.
\end{equation}
A quantization of this curve is to require a basic commutation relation\[[p,x]=g_s\,,\quad p =  g_s \partial_x\,.\] 

A toric Calabi-Yau three-fold can be treated as gluing of various local $\mathbb{C}^3$. Therefore it is not sufficient to know the core region geometry of $\mathbb{C}^3$ without knowing the asymptotic one of it. In \cite{ooguri2000knot, Marino:2004uf}, the Ooguri-Vafa operator actually does the work of gluing core and asymptotic region. The asymptotic region can be obtained by the $S$ transformation of core region geometry. This $S$ transformation is a generator of the modular group $PSL(2,\mathbb{Z})$. Another generator of the modular group is $T$ transformation, which plays a role of framing changing, see \cite{ADKMV} for a detailed analysis.
Acting on canonical doublet $(x,p)$,  $S$ and $T$ have the matrix representation:
\begin{eqnarray}S=\left(\begin{array}{cc} 0&1\\-1&0\end{array}\right)\,\,\,,T=\left(\begin{array}{cc} 1&1\\0&1\end{array}\right)\,.\end{eqnarray} It is easy to check the relation that
\[ST =\left(\begin{array}{cc} 0&1\\-1&-1\end{array}\right)\,,\quad 
(ST)^3 =1\,.\] It is well known that $ST$ transformation generates a $Z_3$ subgroup of $PSL(2,\mathbb{Z})$.

From $S$ transformation we obtain the curve in asymptotic region
\begin{equation}\label{C3right} e^{-x}+e^{p}+1 =0\,.\end{equation} 
However, there are actually three different asymptotic regions of $\mathbb{C}^3$, reflecting the fact that the toric diagram of $\mathbb{C}^3$ has three legs. Therefore the curve (\ref{C3right}) should be triply degenerate. It is easy to check the invariance of (\ref{C3right}) under $ST$. If we denote these three patches as $u, v, w$-patch respectively and define a cyclic relation
\[u=p_w=g_s\partial_w\,,\quad v=p_u=g_s\partial_u\,,\quad w=p_v=g_s\partial_v \,,\] 
 then the $\mathbb{Z}_3$ cyclic symmetry is explicit.
  
From the asymptotic curve in $u$-patch, when $u$ goes to infinity, $v$ should become $i\pi$. It bothers a lot in further analysis. A more convenient way is to throw away the $i\pi$ dependence in all three patches, that is, to reparameterize
\[x\rightarrow x+i\pi, \,\, p\rightarrow p+i\pi\,.\] It changes the core geometry to
\begin{equation}\label{C3core}e^{x}+e^p-1=0\,,\end{equation}
and the asymptotic geometry to
\begin{equation}\label{C3asymptotic}
e^{-x}+e^p-1=0\,.
\end{equation}

Hence the $\mathbb{Z}_3$ symmetry is not generated by $ST$ but by the following $U$-transformation \cite{ADKMV}, 
\begin{equation}
U\left(\begin{array}{c}u\\v\end{array}\right) = ST\left(\begin{array}{c}u\\v\end{array}\right) +\left(\begin{array}{c}0\\i\pi\end{array}\right)\,.
\end{equation}  
It is straightforward to check that $U$ transformation satisfies \[U^3 =1\,.\] 

We claim  core curve (\ref{C3core}) and asymptotic curve (\ref{C3asymptotic}) play important roles in patch-shifting. 

\subsubsection{ $W^3_0$ as the generator of $T$}
Previously we studied  $W$ algebra. However in this article the related symmetry is $W^3_0$, since the local patches are joint by $T$ transformation and $W^3_0$ is the generator of $T$. $T$ acts as follows
\begin{equation}
T:\left(\begin{array}{c}
u\\v
\end{array}\right)=\left(\begin{array}{c}
u+v\\v
\end{array}\right)\,.
\end{equation}
We first notice that $v=g_s\partial_u$ is expressed as $v = g_s\partial_u\varphi(u)$ in a chiral boson theory. Then all arguments follow as we have discussed in previous section. The current related to the transformation is simply $W^3$. Excitated modes of $W^3$ do not contribute because we only consider the asymptotic region  on the $u$-patch (or $v$-patch) as $u\rightarrow\infty$ (or $v\rightarrow\infty$). Therefore only vacuum state contributes otherwise the theory will not be unitary. Thus only $W^3_0$  survives. In summary we conclude that $T$ transformation is generated by $W^3_0$ multiplied by $g_s$.

A $T^n$ transformation of an eigenfunction $f(u)$ has the standard expression
\[e^{ n g_s W^3_0} f(u) e^{- n g_s W^3_0} = f(u+n v)\,.\]

\subsubsection{$S$ transformation on the base}
Now let us consider $S$ transformation on the patches. As we argued before, the $S$ transformation interchanges the canonical pair $(x,p)$ . $x$ and $p$ form a canonical bundle, with the symplectic form as  defined previously. The $S$ transformation also preserves the symplectic structure. If we treat $x$ as the base and $p$ the fiber, $S$ transformation actually bends $x$ to its normal direction. From a physical viewpoint, this can be understood as an insertion of a loop defect which bends the base and fiber simultaneously. Therefore the Hamiltonians on both sides of the defect should  also be related to each other by $S$ transformation.  This is very important in our further analysis.

\section{Zhou\rq{}s Identity and the Topological Vertex}\label{sec:ToV}
In this section, we consider the problem how to obtain partition functions from curves and the underlining symplectic transformations. 

\subsection{Vacuum Partition Function and the Curve of ${\mathbb C}^3$}
Now we look for an eigenfunction of Hamiltonian in the core region
\[H_c(p,x) = e^{x}+e^p-1\,,\]
where the subindex $c$ denotes the core.
  In the local $u$-patch with $u$ being a coordinate on a cylinder, we have asymptotic curve
\[e^{-u}+e^v-1=0\,,\] the Hamiltonian in this region in the complex plane coordinates is 
\begin{equation}
  \label{eq:H_asym}
H_a(L_0,u) = \frac{1}{z_1}+e^{g_s L_0} -1\,,  
\end{equation}
where $L_0 = z_1\partial_{z_1} ,\, z_1 = e^u,\, z_2 = e^v, z_3 =e^w$.
It drives the evolution in the asymptotic region that is $z_1>1$ region, while in $z_1<1$ region, the core curve $H_c$ drives the evolution.

Now we introduce an anti-B-brane at the infinity of $z_1$-plane. The next step is to move it into the core region. This can be treated as an evolution of Hamiltonian. Unfortunately, there are infinitely many evolution paths in the spirit of path integral. Moreover it is quite difficult to approach this problem from a standard Hamiltonian analysis since the Hamiltonian is highly nonlinear.
  
Therefore we need to find a new description of Hamiltonian evolution. Notice the CFT implied by the curve is a Kodaira-Spencer theory, which is equivalent to a free fermion theory. Since a brane (anti-brane) could be understood as a fermion (anti-fermion) insertion in local patch \cite{ADKMV}, the bosonized fermion (anti-fermion) field will have a representation (ignore the zero modes) 
\[e^{\pm\varphi(z)} = \exp\left(\pm\sum_{n\neq 0} \frac{a_{-n}}{n}z^n\right)\,.\] 

It means the evolution of Hamiltonian can be replaced by infinitely many branes insertions in between $z_1=1$ and $z_1=\infty$. This is due to the fact radial ordering on a complex plane is a time ordering on cylinder while the later is controlled by the Hamiltonian. The OPEs of branes and anti-branes now can be understood as propagators.

 However, the positions where B-branes are inserted are arbitrary according to path integral. We may expect a classical equation of motion to determine the orbit completely. However, it is difficult to deal with a quantum system where there are different Hamiltonians in different regions. To simplify the problem a bit in this article we divide the space into the asymptotic region and  the core region associate an Hamiltonian with each coordinate charts. 
We may choose the two coordinate charts to be
\[U_a = \{z_1 = e^u\in(0,\infty]\}\,,\,\,U_c = \{z_1\rq{} = e^{u\rq{}}\in[0,\infty)\}\]
 where $U_a$ is dominated by asymptotic Hamiltonian and $U_c$ is dominated by core Hamiltonian.
 
In $U_a$ chart, we propose the following ansatz equation
\begin{equation}\label{asymptotic}
\langle 0|H_a \exp\left(\sum_{n>0}\frac{a_{n}}{n}z^{-n}\right)\prod_{i\geq 1}^\infty V_-(w_i)|0\rangle\equiv 0\,.
\end{equation}
Applying (\ref{eq:H_asym}) and moving $q^{L_0}$ out of the correlation function, we obtain
\begin{eqnarray*}
&& \langle 0|(q^{L_0} + \frac{1}{z} -1) \exp\left(\sum_{n>0,i\geq 1}\frac{(w_i/z)^n}{n}\right)|0\rangle\\
&=& (q^{L_0}+\frac{1}{z}-1)\prod_{i=1}^\infty (1-w_i/z)^{-1}.
\end{eqnarray*}
The vanishing condition gives rise to
\[\prod_{i=1}^\infty (1-\frac{w_i}{q z})^{-1} = \left(1-\frac{1}{z}\right)\prod_{i=1}^\infty (1-w_i/z)^{-1}\,.\] 
Suppose $w_1 = 1$ and a recursion relation
\[w_{i+1} = q^{-1} w_{i}\,.\] We can prove that it is the solution of the eigen-equation (\ref{asymptotic}). The resulting wave function turns out be a quantum dilogarithm \cite{faddeev1994quantum}, namely
\begin{equation}\label{eigenwave}\Psi^*(u) = \exp \left(\sum_{n>0}\frac{ -q^n e^{-nu}}{n(n)_q}\right)\,,\end{equation} where \[(n)_q = 1-q^n\,.\] 

The orbit of the branes insertions are then a set of discrete points at $\{w_i = q^{-i+1}\}$, $(i\geq 1)$.
This analysis can be generalized to including the evolution of many anti-branes as well. The insertions of these anti-branes can be understood as the generating function for bra Schur states, namely
\[\langle 0|\prod_i V_+^*(z_i) = \sum_\lambda \langle \lambda|s_{\lambda}(z_i)\,.\]
We have chosen anti-brane inserted at infinity. Certainly we can consider brane inserted at infinity where a similar analysis leads to the following ansatz equation
\[\langle 0| H_a V_+(z)\prod_i V_-^*(w_i)|0\rangle =0\,.\] 
Solve the equation we can locate the positions of branes $V_-^*$\rq{}s on the orbit
\[\{w_i = q^{-i+1},\,i\geq 1\}\,.\] 

It is then clear how to determine positions of anti-branes (branes) in $[1,\infty)$. A similar analysis can be done for $U_c$ coordinate chart where the branes insertions are near origin ($e^u=0$). Hence it will locate the orbit points of anti-branes in the region $(0,1]$ by the ansatz equation 
\begin{eqnarray}
&&\langle 0| \prod_{i=1}^\infty V_+^*(w_i)\exp\left(\sum_{n>0}\frac{a_{-n}}{n}z^n\right) H_c|0\rangle=0\,.
\end{eqnarray}
The solution of this equation gives rise to a set of $w_i$\rq{}s
\[\{w_i = q^{i-1},\,\,i\geq 1\}\,.\]

The next step is to join these two coordinate charts into a single $u$-patch as we noted. The anti-brane from infinity and the brane from origin meet at point 1 and annihilate each other identically. 
It actually gives rise to a vacuum partition function in $u$-patch, we get
\begin{eqnarray}\label{vacuumpartition}
&&\langle 0|\prod_{i=1}^\infty V_-(q^{-i+1})q^{L_0}\prod_{j=1}^{\infty}V_+^*(q^{j-1})|0\rangle\\\nonumber &=&\langle 0|\prod_{i=1}^\infty V_-(q^{\rho_i})\prod_{j=1}^{\infty}V_+^*(q^{-\rho_j})|0\rangle=1\,.\end{eqnarray} 
It is what we expect because when we glue two cylinders into a torus, the torus vacuum partition function can be chosen as 1 due to normalization.
However, this result is quite different from the one obtained in \cite{Okounkov:2003sp}, where the vacuum partition function is chosen to be MacMahon function.

There is a subtle feature need to be clarified.  For the vertex operators $V$ and $V^*$, if there are no zero modes, it is not a faithful correspondence between fermion and boson. However  in this article, zero modes will not play significant roles in many calculations. Only if two charts are joined into a single patch with a defect at point 1, must the contribution of zero modes be retrieved. In that case the contribution will highly depend on the representation of the defect. 

It is worth comparing the vertex operator formalism with the definition fermionic vacuum. An observation is that suppose we define a correspondence
\begin{eqnarray}
\label{Dict}
\psi_{\rho_i}\rightarrow V_-(q^{\rho_i})\,,\,\,\psi^*_{-\rho_j}\rightarrow V_+^*(q^{-\rho_i})\,,
\end{eqnarray}
the Dirac sea structure corresponding to fermionic vacuum now becomes
\[\cdots\psi_{-5/2}\psi_{-3/2}
\psi_{-1/2}\psi^*_{1/2}\psi^*_{3/2}\psi^*_{5/2}
\cdots\,.\] Thus it corresponds to inner product of the fermionic vacuum \[\langle vac|vac\rangle\] as we obtained in sec. \ref{sec:vertex} where the fermionic vacuum corresponds to the insertions of branes at infinity and anti-branes at origin. 

The last paragraph is only a rough idea about the projective relation from vertex operators to fermions. We shall have a more concrete derivation of it in next subsection. 

\subsection{Excited States and the Profile of a Young Diagram}
Now we consider excited states in $u$-patch\footnote{for excited states in $v$- or $w$-patch, the same argument follows}. Firstly, suppose there is an excited state labeled by a Young diagram $\lambda$ inserted at infinity of $u$-patch and there are no excitations on the other two patches. The partition function is
\begin{eqnarray}
\langle \lambda|\prod_{i>0} V_-(q^{\rho_i})\prod_{j>0} V_+^*(q^{-\rho_j})|0\rangle = s_{\lambda}(q^{\rho})\,.
\end{eqnarray}

A slightly more complicated case is that besides $\lambda$ there is also an excited state labeled by $\mu$ at the origin of $u$-patch. The the partition function is
\begin{equation}
\langle \lambda|\prod_{i\geq 1}V_-(q^{\rho_i})\prod_{j\geq 1} V_+^*(q^{-\rho_j})|\mu\rangle = \sum_\eta s_{\lambda/\eta}(q^{\rho})s_{\mu/\eta}(q^{\rho})\,.
\end{equation}
To obtain the equality we have used
\begin{eqnarray*}
\sum_\eta \langle \lambda |\prod_j V_-(z_j)|\eta\rangle&=&\sum_{\eta,\xi}\langle 0|s_{\lambda}(a_+)s_{\eta}(a_-)s_{\xi}(a_-)|0\rangle s_\xi(z_j)\\\nonumber
&=&\sum_{\eta,\theta,\xi}c_{\eta\theta}^\lambda \langle \theta|\xi\rangle s_\xi(z_j) = \sum_{\eta,\xi} c_{\eta\xi}^\lambda s_{\xi}(z_j) = \sum_{\eta}s_{\lambda/\eta}(z_j)\,,
\end{eqnarray*}
where $s_{\lambda/\eta}$ is a skew Schur polynomial and $c_{\eta\xi}^\lambda$ is the Littlewood-Richardson coefficient defined by
\[s_{\lambda/\eta} = \sum_{\theta}c_{\eta\theta}^\lambda s_{\theta}\,.\]

Since we argued in previous section, branes and anti-branes can be inserted not only at the infinity of a given patch, but also at the point 1 (more precisely, on the unit circle). Although we have calculated the simple case for excitations near the origin and infinity on a complex plane it would be quite interesting to ask the question about excitations in the bulk near point 1. It corresponds to the case of joining two charts into a single patch with some defect inserted at point 1.

In a local patch, the unit circle does not belong to either the asymptotic region or the core region. Previously we considered the insertion of vacuum at point 1 the fermionic vacuum becomes products of $V_-$ and $V_+^*$\rq{}s, with all $V_-$\rq{}s ($V_+^*$\rq{}s) located to the left (right) side of point 1, namely $V_-$\rq{}s are in the asymptotic region corresponding to the outgoing modes and $V_+^*$\rq{}s are in the core region corresponding to the incoming modes. 

\begin{figure}[h]
  \centering
  \includegraphics[width=0.5\textwidth]{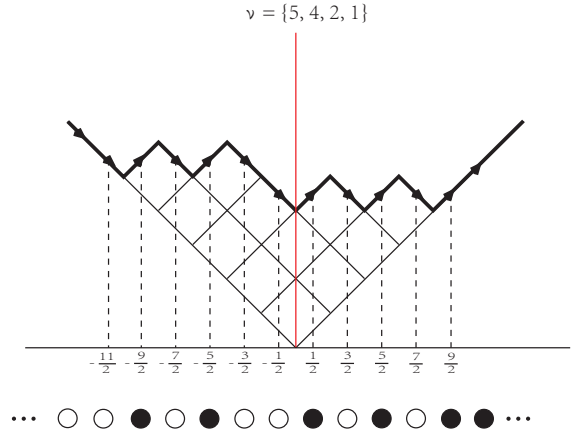}
  \caption{An example of a fermionic excited state and the corresponding Young diagram}
  \label{fig:Profile}
\end{figure}

Now we consider a fermionic excited state labeled by Young diagram $\nu$. In the fermionic picture, it is an excited state from Dirac sea. The $\nu$ state can be written down according to the profile of the Young diagram $\nu$. As in fig. \ref{fig:Profile}, 
where \[\nu = \{5,4,2,1\}\]
the corresponding fermionic excitations are
\[\cdots\psi_{-11/2}\psi^*_{-9/2}\psi_{-7/2}
\psi^*_{-5/2}\psi_{-3/2}\psi_{-1/2}\psi^*_{1/2}
\psi_{3/2}\psi^*_{5/2}\psi_{7/2}\psi^*_{9/2}
\cdots\,\,.\]

For a general $\nu$, the modes of $\psi$ (white dots) belong to the set
\begin{equation}\label{white dots}
\left\{\cdots,\,\,\nu^t_3 -3+\frac{1}{2}\,,\,\nu^t_2-2+\frac{1}{2},\,\,\nu^t_1-1+\frac{1}{2}\right\} \equiv \{\nu^t+\rho\}\,.
\end{equation}
Similarly the modes of $\psi^*$ (black dots) belong to the set
\begin{equation}\label{white dot}
\left\{-\nu_1+1-\frac{1}{2},\,\,-\nu_2+2-\frac{1}{2},\,\,-\nu_3 +3-\frac{1}{2},\,\cdots\right\} \equiv \{-\nu-\rho\}\,.
\end{equation}

In this fermionic picture, it is clear that presumably, there is an infinity height fermionic tower at point $e^u=1$. This tower will expand to elsewhere in $u$-patch due to quantum shift.

For the case $\nu=\phi$, the empty set, we have already seen this quantum shift changes the vacuum Dirac sea to an infinite products of $V$ and $V^*$\rq{}s. Actually, it is very simple to deduce from the curve. At point $z_1=1$, the Hamiltonian just becomes 
\[H(L_0,1) = q^{L_0}\,.\] 
The fermionic modes expansion becomes
\[\psi(1) = \sum_{r\in\mathbb{Z}-\frac12}\psi_r\,,\,\,\psi^*(1) = \sum_{r\in\mathbb{Z}-\frac12}\psi^*_r\,.\]
In a quantum manner, all excitations are including in multi-products of these fields. For the physical vacuum $vac$, it is a  multi-product in sequence as \[\cdots\psi_{-5/2}\psi_{-3/2}\psi_{-1/2}
\psi^*_{1/2}\psi^*_{3/2}\psi^*_{5/2}\cdots\,\,.\] 

 Hamiltonian at point 1 is a transport operator moving all $\psi$- fields to the left of 1 and $\psi^*$-fields to the right of 1. Further according to the bosonization formula, we reproduce the vacuum partition function as 
\begin{eqnarray}\label{vactower}
\langle0|\prod(q^{L_0}V_-(1))q^{L_0}\prod(V_+^*(1)q^{L_0})|0\rangle\,.
\end{eqnarray} It is just another expression of (\ref{vacuumpartition}). Here the left (right) transporting behavior is transferred to left (right) action on the vertex operators. It proves the projective relation as we mentioned in eq. (\ref{Dict}). 

If we want to generalize the analysis to a generic $\nu$ state, then we just need to reshuffle (\ref{vactower}) according to the profile \footnote{here profile means the sequence of $V_-$ and $V^*_+$'s is determined according to the profile by the projective relation} of the Young diagram of $\nu$. Hence it gives rise to
\begin{eqnarray}\label{profile}
\langle0|\prod_{\text{profile}\,\,\nu}V_-(q^{\nu^t+\rho})V_+^*(q^{-\nu-\rho})|0\rangle\,.
\end{eqnarray}
Moving all $V_+^*$\rq{}s to the right side of all $V_-$\rq{}s we get
\begin{eqnarray}&\langle0&\hspace{-2mm}|\prod_{\text{profile}\,\,\nu}V_-(q^{\nu^t+\rho})V_+^*(q^{-\nu-\rho})|0\rangle\\\nonumber
&=& \prod_{(i,j)\in\nu}\frac{1}{1-q^{h(i,j)}}\equiv Z_{\nu}(q)\,,\end{eqnarray} 
where \begin{eqnarray}Z_\nu(q) &:= &\prod_{i,j\in\nu}\frac{1}{1-q^{h(i,j)}} = Z_{\nu^t}(q)\nonumber\\
&=&(-)^{|\nu|}\prod_{(i,j)\in\nu}\frac{q^{-h(i,j)}}{1-q^{-h(i,j)}}\\\nonumber&=&(-)^{|\nu|}q^{-||\nu||/2-||\nu^t||/2}\prod_{(i,j\in\nu)}\frac{1}{1-q^{-h(i,j)}}\,,\end{eqnarray} with $h(i,j)$ being the hook length of square $(i,j)$ in $\nu$.
Notice that $Z_\nu$ is neither $s_{\nu}(q^{-\rho})$ nor $s_{\nu^t}(q^{-\rho})$. Schur polynomial in variables $\{q^{1/2},q^{3/2},q^{5/2},\cdots\}$ is
\begin{eqnarray}
s_{\nu}(q^{-\rho}) = q^{\frac{||\nu^t||}{2}}\prod_{i,j\in\nu}\frac{1}{1-q^{h(i,j)}}= (-)^{|\nu|}s_{\nu^t}(q^{\rho})\,.
\end{eqnarray}
 
Now we consider $V_-$ and $V^*_+$'s insertions respectively. For the $V_-$\rq{}s insertions, we have
\begin{eqnarray}\label{brainsertion}
\langle 0| \prod_{i\geq 1} V_-(q^{\nu^t+\rho_i})\,.
\end{eqnarray}
A Young diagram $\nu$ in terms of Frobenius notation is \[\{r_1,r_2,\cdots,r_d|s_1,s_2,\cdots,s_d\}\] where 
\[r_i = \nu_i-i+\frac{1}{2},\,\,s_i = \nu^t_i-i+\frac{1}{2}\,.\]
According to the projective relation the fermionic bra state can be represented as
\begin{eqnarray}
&\langle&\hspace{-3.5mm} \Omega|\cdots\psi_{\nu^t_i-i+\frac{1}{2}}\psi_{\nu^t_{i-1}-i+\frac{3}{2}}\cdots\psi_{\nu^t_1-\frac{1}{2}}\\\nonumber
&=&\langle vac|(-)^{\sum_{i=1}^{d}(s_i-\frac{1}{2})}\prod_i^d\psi_{s_i}\psi^*_{r_i}=\langle \nu|\,.
\end{eqnarray}
Similarly, for $V_-$\rq{}s insertions, the corresponding fermionic ket state is 
\begin{equation}(-)^{\sum_i^d(r_i-\frac{1}{2})}\prod_i^d\psi_{-s_i}\psi^*_{-r_i}|vac\rangle = |\nu^t\rangle\,.\end{equation}
The states are compatible with the geometrical observation from infinity to the origin on one local patch. The $S$ transformation which exchanges canonical variables (position and momentum) \lq\lq{}bends\rq\rq{} the project line to its normal at point 1. Then near infinity, we see the profile of $\nu$, while near the  origin, we find that it reflects to $\nu^t$.

This observation defines the following rules:

1. From infinity to 1, the representation has not been changed.

2. From 1 to 0, the representation becomes its transpose. 

In summary we can consider the patch-shifting and its impacts on the vertex operator formalism.

We propose a configuration 
\begin{eqnarray}\label{OurVertex}
\langle \lambda,\nu,\mu\rangle& \equiv&(-)^{|\nu|}q^{\frac{||\nu||}{2}}\langle \lambda |\prod_{\text{profile}\,\, \nu}V_-(q^{\nu^t+\rho})V_+^*(q^{-\nu-\rho})|\mu\rangle\\\nonumber
&=& s_{\nu}(q^{\rho})\sum_{\eta}s_{\lambda/\eta}(q^{\nu^t+\rho})s_{\mu/\eta}(q^{\nu+\rho}).
\end{eqnarray} 
The factor $(-)^{|\nu|} q^{||\nu||/2}$ comes from zero modes of $V$ and $V^*$. Actually, if we keep the Boson-Fermion correspondence being exact, we should include the contribution of zero modes. The result of normal ordering now becomes:
\begin{eqnarray}
\label{eq:zeronorm}
\prod_{(i,j)\in\nu}\frac{1}{q^{-\nu_i-\rho_i}-q^{\nu^t_j+\rho_j}}&=&(-)^{|\nu|}q^{||\nu||/2-||\nu^t||}\prod_{(i,j)\in\nu}\frac{1}{1-q^{-h(i,j)}}\\\nonumber&=(-)^{|\nu|}&q^{\kappa_{\nu}/2}s_{\nu}(q^{\rho})\,.
\end{eqnarray}
Then up to a framing factor $(-)^{|\nu|}q^{\kappa_\nu/2}$, the Schur function $s_{\nu}(q^\rho)$ occurs as desired.

The states under the shifting from a $u$-patch to a $v$-patch are compatible with the corresponding curves of different charts on patches.
 
 For example, the insertion of the bra state $\lambda$ at infinity on $u$-patch  is an insertion at point 1 in $v$-patch. Thus patch-shifting leads to bringing a $\lambda$ state from infinity of $u$ to 1 of $v$.
 
   Then a $\nu$ insertion at point 1 in the $u$-patch becomes a ket state $\nu^t$ inserted at the core region in $v$-patch.
   
     Similarly a ket state $\mu$ inserted in the core region determined by
\[e^{u}+e^v-1 = 0\] in the $u$-patch should be transformed to the asymptotic region in $v$-patch by $S$-transformation, and the $T$ transformation is required to cancel the divergence. For example 
  \[e^{-u-v}+e^{-v}-1=0\] as $u$ goes to $-\infty$, $v$ becomes $\infty$, this operation moves $\mu$ ket state to a $\mu^t$ bra state along with a factor
  $q^{\kappa_{\mu}/2}$ due to the $T$ transformation.
  
To join the asymptotic region and the core region together into a T-transformed $v$-patch, we need $T$-transform the core region (with $\nu^t$ inserted on) and also the defect (representation $\lambda$). It results in a further $q^{\kappa_{\nu}/2}$ factor in the expression in $v$-patch. Notice that there is no further factor corresponding to a $\lambda$ insertion at point 1 since \[q^{\kappa_{\lambda}/2}q^{\kappa_{\lambda^t}/2}=1\,.\]
  
  Now we have the following conjecture
  \begin{eqnarray}\label{Conjecture}
  \langle \lambda, \nu,\mu\rangle &=& q^{\frac{\kappa_{\mu}+\kappa_{\nu}}{2}}\langle \mu^t ,\lambda,\nu^t\rangle\\\nonumber
  &=&q^{\frac{\kappa_{\lambda}+\kappa_{\mu}}{2}}\langle \nu,\mu^t,\lambda^t\rangle\,.
  \end{eqnarray}
It is our major observation from the curve of $\mathbb{C}^3$.

It is difficult to verify this conjecture directly. However, if we let one of the representations $\lambda$, $\mu$ and $\nu$ be an empty representation $\phi\equiv 0$, then the resulting identities are just Zhou\rq{}s identities \cite{zhou2003conjecture}. 

For example, let $\nu =0$. We have 
\begin{eqnarray}
\langle \lambda,0,\mu\rangle &=& \sum_{\eta}s_{\lambda/\eta}(q^{\rho})s_{\mu/\eta}(q^{\rho})\\\nonumber
&=&q^{\kappa_{\mu}/2}\langle \mu^t, \lambda, 0\rangle = q^{\kappa_{\mu}/2}s_{\lambda}(q^{\rho})s_{\mu^t}(q^{\lambda^t+\rho})\\\nonumber
&=& q^{(\kappa_{\lambda}+\kappa_{\mu})/2}\langle 0,\mu^t,\lambda^t\rangle
\\\nonumber
&=& q^{(\kappa_{\lambda}+\kappa_{\mu})/2}s_{\mu^t}(q^{\rho})s_{\lambda^t}(q^{\mu^t+\rho})\,.
\end{eqnarray}
It is nothing but Zhou\rq{}s identity.

We can verify other degenerate cases of (\ref{Conjecture}) in detail. Consequently we get Zhou\rq{}s identities in all cases.

\subsection{The relation with the Topological Vetex}
It would be interesting to compare eq. (\ref{OurVertex}) with the famous topological vertex  proposed in \cite{top_vertex} and further the topological vertex in terms of symmetric polynomials in \cite{Okounkov:2003sp} and \cite{zhou2003conjecture}.

The topological vertex in \cite{Okounkov:2003sp, zhou2003conjecture} is defined as
\begin{equation}\label{OV2003}
C(\lambda,\,\mu,\,\nu) = q^{\kappa_\lambda/2}s_{\nu}(q^\rho)\sum_{\eta}s_{\mu/\eta}(q^{\nu^t+\rho})s_{\lambda^t/\eta}(q^{\nu+\rho})
\end{equation}
In our configuration 
\begin{equation}
C(\lambda,\,\mu,\,\nu)=q^{\kappa_{\lambda}/2}\langle \mu,\,\nu,\,\lambda^t\rangle\,.
\end{equation}
It means what we have obtained is a reformulation of the topological vertex. However, the approach here is quite different from that in \cite{top_vertex} and \cite{Okounkov:2003sp}. An direct observation is that our definition as in eq. (\ref{OurVertex}) has a very clear patch meaning rather than a unified topological vertex. The cyclic symmetry of the topological vertex now becomes the shifting of patches.

\section{Conclusions}\label{sec:conclusion}

We find an explicit correspondence between A- and B-model for the case of topological vertex. In our opinion, the mirror curve of $\mathbb{C}^3$ is not a global ly defined chart but a union of two coordinate charts within defects inserting at point 1. It is crucial for deriving B-model correlation function, which becomes A-model topological invariant. A new vertex operator approach to the topological vertex is proposed. On the way of doing this, we prove the conjecture proposed in \cite{ADKMV}. The vertex operator approach can be treated as an application of projective representation introduced in \cite{okounkov2001infinite}. Finally, we propose a conjecture on the topological vertex (or in B-model, a three-leg correlation function) identity (\ref{Conjecture}), which becomes Zhou\rq{}s identities of Hopf links in degenerate cases.

There are many further works in this direction. We just list three of them for instance. Firstly, the identity (\ref{Conjecture}) is new and a mathematical proof is not known to the authors. Secondly, the vertex operator approach could be generalized to other curves associated to many toric Calabi-Yau manifolds. Due to the identity (\ref{Conjecture}), it is quite free to glue topological vertices to formulate complicated toric Calabi-Yau\rq{}s. This calculation is working in progress and a future article will contain some applications. Thirdly, it is natural to ask for a refined version of this approach. However, this is quite difficult since there the refined curve \footnote{Eynard and Kozcaz provided a mirror curve for refine topological vertex in \cite{Eynard:2011vs}, the curve has no simple expression as the topological vertex.} is very complicated and related symplectic transformations are not well-known. Maybe a simpler case could be considered first. For example, when a background charge is introduced into the Kodaira-Spencer theory the resulting theory is hence the Feign-Fuchs bosonic theory. The underlining integrability is controlled by the Calogero-Sutherland model \cite{calogero2003solution, sutherland1971exact}. In this case, two refined parameters($t$ and $q$) are related by $t=q^{\alpha}$ (twisted case) and the eigenfunctions are Jack symmetric functions in the limit $q\rightarrow 1$. A very similar analysis could be done for this case. We expect a Jack symmetric function expression for the twisted topological vertex. 

\section*{Acknowledgments}
We would like to thank Professor Guoce Xin, Professor Ming Yu and Professor Jian Zhou  for valuable comments.
The authors are grateful to Morningside Center of Chinese Academy of Sciences
and Kavli Institute for Theoretical Physics China at the Chinese Academy of
Sciences for providing excellent research environment. This work is also partially supported by Beijing Municipal Education Commission Foundation (KZ201210028032, KM201210028006), Beijing Outstanding Person Training Funding (2013A005016000003).

\appendix

\section{Some Notations on Symmetric Polynomials}
In this appendix we just provide a brief review of some symmetric functions. For detailed description please look up the book by Macdonald \cite{Macdonald1999}.
\begin{Def}
  An {\bf elementary symmetric polynomial} is defined by
  \begin{equation}
    \label{eq:elem}
    e_r(x_1, x_2, \cdots)=\sum_{i_1<i_2<\cdots <i_r}x_{i_1}x_{i_2}\cdots x_{i_r},
  \end{equation}
for $r\geq 1$ and $e_0=1$. 
\end{Def}
The generating function for the $e_r$ is
\begin{equation*}
  E(t)=\sum_{r\geq 0}e_rt^r=\prod_{i\geq 1}(1+x_it).
\end{equation*}

\begin{Def}
  A {\bf complete (homogenous) symmetric polynomial} is defined by
  \begin{equation}
    \label{eq:comp}
    h_r(x_1, x_2, \cdots)=\sum_{i_1\leq i_2\leq\cdots\leq i_r}x_{i_1}x_{i_2}\cdots x_{i_r},
  \end{equation}
for $r\geq 1$ and $h_0=1$.
\end{Def}
The generating function for the $h_r$ is
  \begin{equation*}
    H(t)=\sum_{r\geq 0}h_rt^r=\prod_{i\geq 1}\frac{1}{1-x_it}.
  \end{equation*}
\begin{Def}
A {\bf Schur polynomial} $s_\lambda$ as a symmetric polynomial in variables $x_1, x_2, \cdots$ corresponding to a partition $\lambda$ is defined by
\begin{equation}
  \label{eq:schur}
s_{\lambda}(x_1,\cdots,x_N):= \sum_{T}\bold{x}^T  
\end{equation}
where $T$ is a semi-standard tableau of shape $\lambda$ and $\bold{x}^T
=\prod_ix_i^{n_i}$ with $n_i$ the number of $i$ filling in $T$.
\end{Def}

\begin{Def}[Jacobi-Trudi]
The Schur polynomial can be calculated from the elementary or complete polynomials by
\begin{equation}
\label{eq:Jacobi-Trudi}
  s_\nu(x_1, x_2, \cdots, x_n)=\det (h_{\nu_i-i+j})=\det (e_{\nu^t_i-i+j}).
\end{equation}
\end{Def}
Now suppose the variables ($x_1, x_2, \cdots $) appear in a formal power series $E(t)=\prod_i(1+x_it)$. We simply denote the Schur function by
\begin{equation*}
  s_\nu(E(t)).
\end{equation*}

For example
\begin{equation}
  E(t)=\prod_{i=0}^\infty (1+q^it)=\sum_{r=0}^\infty e_rt^r
\end{equation}
where
\begin{equation}
\label{eq:elem_for_q}
e_r = \prod_{i=1}^r\frac{q^{i-1}}{1-q^i}.
\end{equation}
Hence the corresponding Schur function is written as $s_\lambda(1, q, q^2, \cdots)$. In the q-number notation $[x]=q^{x/2}-q^{-x/2}$
\begin{equation*}
s_\nu(q^{-\rho})  =(-1)^{|\nu|}q^{-\kappa(\nu)/4}\prod_{x\in\nu}\frac{1}{[h(x)]}
\end{equation*}
where $h(x)$ is the hook length of the square $x$ and $\kappa(\nu)=2(n(\nu^t)-n(\nu))$ with $n(\nu)=\sum_i \nu_i(i-1)$.

Now let us generalize the formal power series to a more complicated case
\begin{equation}
  \label{eq:E_mu}
  E_\mu(t)=\prod_{i=1}^\infty(1+q^{\mu_i-i+1/2}t)=\prod_{i=1}^\ell \frac{1+q^{\mu_i-i+1/2}t}{1+q^{-i+1/2}t}\prod_{i=1}^\infty (1+q^{-i+1/2}t).
\end{equation}
Recall a very useful identification between multisets of number
\begin{equation}
  \label{eq:set}
  \{\mu_i-i, (d<i\leq \ell)\}=\{-1,\cdots, -\ell\}-\{-\mu^t_i+i-1, (1\leq i\leq d)\}
\end{equation}
where $d$ is the diagonal of $\nu$. 
According to Frobenius notation $\nu=(\alpha_1, \cdots, \alpha_d|\beta_1,\cdots, \beta_d)$, it can be written as 
\begin{equation}
    \{\mu_i-i, (d<i\leq \ell)\}=\{-1,\cdots, -\ell\}-\{-\beta_i-1, (1\leq i\leq d)\}.
\end{equation}
 (\ref{eq:E_mu}) becomes
 \begin{equation}
   \label{eq:E_mu_Frob}
   E_\mu(t)=\prod_{i=1}^{d(\mu)}\frac{1+q^{\alpha_i+1/2}t}{1+q^{-\beta_i-1/2}t}\prod_{i=1}^\infty (1+q^{-i+1/2}t).
 \end{equation}
Therefore
 \begin{equation}
   \label{eq:Schur_link}
   s_\nu(E_\mu(t))=s_\nu(q^{\mu_1-1+1/2}, q^{\mu_2-2+1/2}, \cdots)\,,
 \end{equation}
or it can be put in a simple notation $s_\nu(q^{\mu+\rho})$ where $\rho=-\frac12, -\frac32, \cdots$. In the Frobenius notation
\begin{equation}
  \label{eq:Schur_link_Frob}
  s_\nu(E_\mu(t))=s_\nu(q^{\alpha_1+\frac12}, \cdots, q^{\alpha_{d(\mu)}+\frac12}, q^{-\frac12}, \cdots, \widehat{q^{-\beta_1-\frac12}}, \cdots, \widehat{q^{-\beta_{d(\mu)}-\frac12}}, q^{-\beta_{d(\mu)}-\frac32},\cdots ).
\end{equation}

\begin{Def}
  A {\bf skew Schur polynomial} $s_{\lambda/\mu}$ as a symmetric function in variables $x_1, x_2, \cdots$ is defined by
  \begin{equation}
    \label{eq:skew_schur}
    s_{\lambda/\mu}(x_1, x_2, \cdots)=\sum_{T} \bold{x}^T
  \end{equation}
where $T$ is a semi-standard tableau of shape $\lambda-\mu$. 
\end{Def}
The skew Schur function has a property
\begin{equation*}
  s_{\lambda/\mu}(x, y)=\sum_\nu s_{\lambda/\nu}(x)s_{\nu/\mu}(y).
\end{equation*}
Therefore it can be generalized to $n$ sets of variables $x^{(1)}, \cdots, x^{(n)}$
\begin{equation}
  s_{\lambda/\mu}(x^{(1)}, \cdots, x^{(n)})=\sum_{(\nu)}\prod_{i=1} s_{\nu^{(i)}/\nu^{(i-1)}}(x^{(i)})
\end{equation}
summed over all sequences $(\nu)=(\nu^{(0)}, \cdots, \nu^{(n)})$ of partitions such that $\nu^{(0)}=\mu$, $\nu^{(n)}=\lambda$, and $\nu^{(0)}\subset\cdots\subset\nu^{(n)}$.

\begin{Def}[Jacobi-Trudi]
The skew Schur polynomial also can be calculated from the elementary or complete polynomials by
\begin{equation}
\label{eq:skew_Jacobi-Trudi}
  s_{\lambda/\mu}(x_1, x_2, \cdots, x_n)=\det (h_{\lambda_i-\mu_j-i+j})=\det (e_{\lambda_i^t-\mu^t_j-i+j}).
\end{equation}
\end{Def}

\section{ The identity}\label{app:id}
In this appendix we provide a combinatoric proof of the identity
 \begin{equation}
  s_{\lambda /\mu}( q^{\nu +\rho}) = (-1)^{|\lambda|-|\mu|} 
s_{\lambda^t /\mu^t}( q^{-\nu^t -\rho}).
 \end{equation}
According to the definition of $s_{\lambda/\mu}$ (\ref{eq:skew_Jacobi-Trudi}) we only need to prove
\begin{eqnarray*}
  h_r(q^{\nu+\rho})=(-1)^re_r(q^{-\nu^t-\rho}).
\end{eqnarray*}
\begin{proof}
Now we use the Frobenius notation of a partition $\nu=(\alpha_1, \cdots, \alpha_d|\beta_1,\cdots, \beta_d)$.
Suppose $\nu_1=N$, $\nu_1^t=k$
  \begin{eqnarray}
    E(t, q^{-(\nu^t+\rho)})
&=&(1+q^{-(\nu_1^t-1+1/2)}t)\cdots (1+q^{-(\nu^t_j-j+1/2)}t)\cdots (1+q^{-(\nu^t_N-N+1/2)}t)\times\nonumber\\
&&\times(1+q^{-(-(N+1)+1/2)}t)\cdots \nonumber\\
&=&\prod_{i=1}^d\frac{1+q^{-(\beta_i+1/2)}t}{1+q^{\alpha_i+1/2}t}
E_0(t)\label{eq:elementary}
  \end{eqnarray}
where $E_0(t)=\prod (1+q^{-\rho}t)$.
We have used an identity among multisets of number
\begin{equation*}
  \{1, 2, \cdots, N\}=\{ j-\nu^t_j, (N\geq j>d)\}\cup \{\alpha_i+1 (i=1, \cdots, d) \}.
\end{equation*}
Similarly 
\begin{eqnarray}
  H(t, q^{\nu+\rho})
&=&\frac{1}{1-q^{\nu_1-1+1/2}t}\cdots\frac{1}{1-q^{\nu_i-i+1/2}t}\cdots
\frac{1}{1-q^{\nu_k-k+1/2}t}\frac{1}{1-q^{-(k+1)+1/2}t}\nonumber \\
&=&\prod_{i=1}^d\frac{1-q^{-(\beta_i+1/2)}t}{1-q^{\alpha_i+1/2}t}H_0(t)
\label{eq:complete}
\end{eqnarray}
where $H_0(t)=\prod(1-q^{-\rho}t)^{-1}$
and 
\begin{equation*}
  \{1, 2, \cdots, k\}=\{ i-\nu_i, (k\geq i>d)\}\cup \{\beta_i+1 (i=1, \cdots, d) \}.
\end{equation*}

The first factor in (\ref{eq:elementary}) and (\ref{eq:complete}) are almost the same except the $+$ and $-$ sign in front of $q$. In addition $\prod (1-q^{-\rho} t)$ and $\prod(1-q^{\rho}t)^{-1}$ have the same power expansion of $t$.
The difference can be resolved by 
\begin{eqnarray*}
  E(-t, q^{-\nu^t-\rho})=H(t, q^{\nu+\rho}).
\end{eqnarray*}
Therefore we obtain the result we want
\begin{eqnarray*}
  e_r(q^{-\nu^t-\rho})=(-1)^rh_r(q^{\nu+\rho}).
\end{eqnarray*}

\end{proof}

\end{document}